\documentclass[prl,aps,twocolumn,epsfig,superscriptaddress,showpacs]{revtex4}
\usepackage{bm}
\usepackage{amsfonts}
\usepackage{mathrsfs}
\usepackage[intlimits]{amsmath}
\usepackage[colorlinks, citecolor=red]{hyperref}
\usepackage[dvips]{graphicx}

\newcommand{\dsum}{\displaystyle\sum}

\begin{document}

\title{Fault Tolerant Quantum Random Number Generator Certified by Majorana
Fermions}
\author{Dong-Ling Deng}
\affiliation{Department of Physics, University of Michigan, Ann Arbor, Michigan 48109, USA}
\affiliation{Center for Quantum Information, IIIS, Tsinghua University, Beijing, China}
\author{Lu-Ming Duan}
\email{lmduan@umich.edu}
\affiliation{Department of Physics, University of Michigan, Ann Arbor, Michigan 48109, USA}
\affiliation{Center for Quantum Information, IIIS, Tsinghua University, Beijing, China}
\date{\today}

\begin{abstract}
Braiding of Majorana fermions gives accurate
topological quantum operations that are intrinsically robust to noise and
imperfection, providing a natural method to realize fault-tolerant quantum
information processing. Unfortunately, it is known that
braiding of Majorana fermions is not sufficient for implementation of
universal quantum computation. Here we show that
topological manipulation of Majorana fermions provides the full set of
operations required to generate random numbers by way of quantum mechanics
and to certify its genuine randomness through violation of a multipartite Bell inequality. The result opens a new perspective
to apply Majorana fermions for robust generation of certified random
numbers, which has important applications in cryptography and other related
areas.
\end{abstract}


\maketitle


The complex-valued solutions to the Dirac equation predict that every
elementary particle should have a complex conjugate counterpart, namely an
antiparticle. For example, an electron has a positron as its antiparticle.
However, in $1937$ Ettore Majorana~\cite{1937Majorana} showed that the
complex Dirac equation can be modified to permit real wave-functions,
leading to the possible existence of the so called \textquotedblleft \textit{%
Majorana fermions}\textquotedblright\ which are their own antiparticles~\cite{2009Wilczek}. In
condensed matter physics, Majorana fermions may appear as elementary
qusi-particle excitations. To search for Majorana fermions, several
proposals have been made in recent years, including $\nu =5/2$ fractional
quantum Hall system~\cite{2010Stern,2008Nayak}, topological insulator
(TI)---superconductor (SC) interface~\cite{2008Fu}, interacting quantum
spins~\cite{2006Kitaev}, chiral p-wave superconductors~\cite{2000Read},
spin-orbit coupled semiconductor thin film~\cite{2010Sau} or quantum
nanowire~\cite{2010Lutchyn,2010Oreg} in the proximity of an external s-wave
superconductor. Based on these proposals, experimentalists have made great
progress recently. For instance, Ref.~\cite{2012Wang} reported an
experimental observation of coexistence of the superconducting gap and the
topological surface state in the $\mathtt{Bi}_{2}\mathtt{Se}_{3}$ thin film
as a step towards realization of Majorana fermions. More recently, signature
of Majorana fermions in hybrid superconductor-semiconductor nanowire device
has been reported \cite{2012Mourik}, which has raised strong interest in the
community.

Majorana fermions are exotic particles classified as non-abelian anyons with
fractional statistics, and braiding between them gives nontrivial quantum
operations that are topological in nature. These topological quantum
operations are intrinsically robust to noise and experimental imperfection,
so they provide a natural solution to realization of fault-tolerant quantum
gates. Application of Majorana fermions in implementation of fault-tolerant
quantum computation has raised great interest~\cite{2006Kitaev,2008Nayak}.
Unfortunately, braiding of Majorana fermions are not sufficient yet for
realization of universal quantum computation \cite{2008Nayak}, and we need
assistance from additional non-topological quantum gates which are prone to
influence of noise.

In this Letter, we show that topological manipulation of Majorana fermions alone can
be used to realize a quantum random number generator in a fault tolerant
fashion and to certifies its genuine randomness through violation of the
Mermin-Ardehali-Belinskii-Klyshko (MABK) inequality \cite%
{1993Belinskii,1992Ardehali,1990Mermin}. Random numbers have tremendous
applications in science and engineering \cite%
{2001Ackermann,1991Hultquist,2003Tu}. However, generation of genuine random
numbers is a challenging task~\cite{2010Pironio}. Any classical device does
not generate genuine randomness as it allows a deterministic description in
principle. Quantum mechanics is intrinsically random, and one can explore
this feature to generate random numbers \cite%
{1956Isida,1990Svozil,1994Rarity,2000Jennewein}. However, in real
experiments, the intrinsic randomness of quantum mechanics is always
mixed-up with an apparent randomness due to noise or imperfect control of
the experiment~\cite{2010Pironio}. The latter can be exploited by an
adversary opponent and leads to security loopholes in various applications
of randomness. Recently, a nice idea has been put forward to certify genuine
randomness generated by a quantum device through test of violation of the
Bell-CHSH (Clauser-Horn- Shimony-Holt \cite{1969Clauser}) inequality~\cite%
{2007Colbeck,2010Pironio}, and the idea has been demonstrated in a
proof-of-principle experiment using remote entangled ions \cite{2010Pironio}%
. This implementation is not fault-tolerant yet as the remote entanglement
is sensitive to noise and the quantum gates have limited precision which can
all lead to security loopholes. We show here that all the operations for
generation and certification of genuine randomness can be realized through
topological manipulation of Majorana fermions. This implementation is
inherently fault-tolerant and automatically closes security loopholes caused
by influence of noise.

The implementation of certification of a quantum random number generator
with Majorana fermions is tricky. First of all, one can not use the
Bell-CHSH inequality anymore as proposed in Ref. \cite{2010Pironio}, since
it is impossible to violate this inequality through topological manipulation
of Majorana fermions alone \cite{2009Brennen}. In fact,  to observe violations of the CHSH inequality, measurements in the non-Clifford bases are required. However, topological operations on Majorana fermions can only give gates in the Clifford group, and thus not able to achieve the measurements required for the CHSH inequality violation for randomness certification. Consequently, we have to consider
certification of randomness based on extension of the Bell inequalities in
the multi-qubit case. For simplicity, here we use the MABK inequality~for
three logical qubits \cite{1993Belinskii,1992Ardehali,1990Mermin}. We show
that first, this inequality can be used to certify randomness, and second,
the inequality can be tested with topological manipulation of Majorana
fermions alone. For the MABK inequality, we consider three qubits, each with
two measurement settings. We denote the measurement settings for each qubit
by the binary variables $x$, $y$, $z$, and the corresponding measurement
outcomes by $a$, $b$, $c$, where $x,y,z,a,b,c=0,1$. The MABK inequality can
be rewritten as~\cite{1993Belinskii,1992Ardehali,1990Mermin}
\begin{equation}
L\equiv \sum_{(x,y,z)\in \mathcal{S}}\tau (x,y,z)[P(\mathtt{even}|xyz)-P(%
\mathtt{odd}|xyz)]\leq 2,  \label{MABK-Ineq}
\end{equation}%
where $\mathcal{S}=\{(0,0,0),(0,1,1),(1,0,1),(1,1,0)\}$ and $\tau (x,y,z)$
is a sign function defined by $\tau (x,y,z)=(-1)^{(x+y+z)/2}$; $P(\mathtt{%
even}|xyz)$ ($P(\mathtt{even}|xyz)$) is the probability that $a+b+c$ is an
even (odd) number when settings $(x,y,z)$ are chosen. The inequality~(\ref%
{MABK-Ineq}) is satisfied by all local hidden variable models. However, in
quantum mechanics certain measurements performed on entangled states can
violate this inequality. Experimentally, we can repeat the experiment $k$
times in succession to estimate the violation. For each trial, the
measurement choices $(x,y,z)$ are generated by an independent identical
probability distribution $P(xyz)$. Denote the input string as $\mathcal{I}%
=(x_{1},y_{1},z_{1};\cdots ;x_{k},y_{k},z_{k})$ and the corresponding output
string as $\mathcal{O}=(a_{1},b_{1},c_{1};\cdots ;a_{k},b_{k},c_{k})$. The
estimated violation of the MABK\ inequality can be obtained from the
observed data as
\begin{equation}
\hat{L}=\frac{1}{k}\sum_{(x,y,z)\in \mathcal{S}}\frac{\tau (x,y,z)}{P(xyz)}%
[N(\mathtt{even}|xyz)-N(\mathtt{odd}|xyz)],
\end{equation}%
where $N(\mathtt{even}|xyz)$ ($N(\mathtt{odd}|xyz)$) denotes the number of
trials that we get an even (odd) outcome $a+b+c$ after $k$ times of
measurements with the measurement setting $(x,y,z)$.

\begin{figure}[tbp]
\includegraphics[width=88mm]{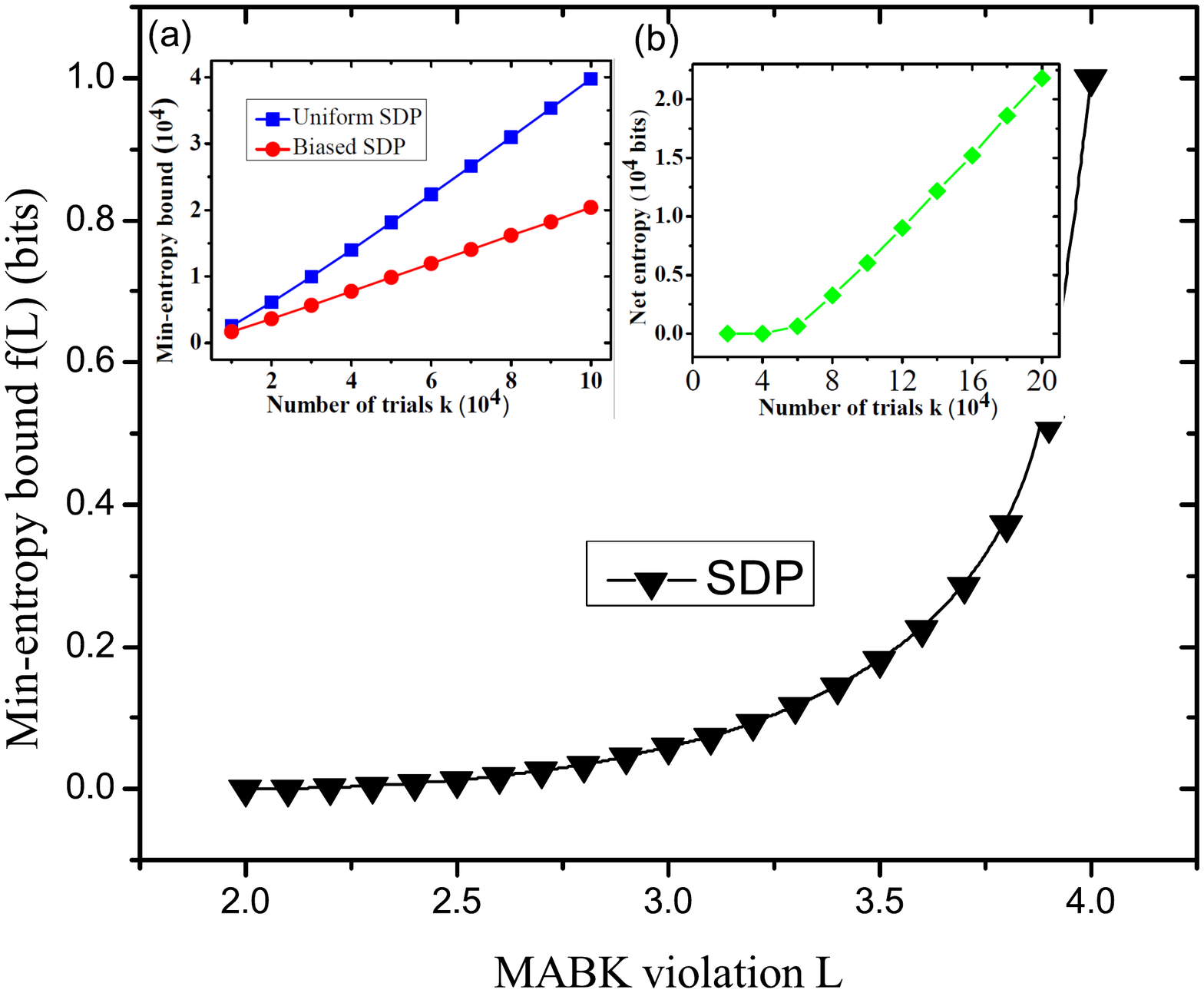}\newline
\caption{(Color online) Plot of the function $f(\hat{L})$ versus violation $%
\hat{L}$ of the MABK inequality. The function is calculated through
optimization based on the semi-definite programming with the details shown
in the Supplemental Material. The inset (a) shows the lower bound of the
min-entropy $kf(\mathcal{L}_m-\protect\epsilon)-\log_2\frac{1}{\protect\delta%
}$ versus the number of trials $k$. Here we assume an observed MABK
violation lies within the interval $3.9=\mathcal{L}_m\leq \hat{L}<\mathcal{L}%
_{max}=4$ with probability $\protect\delta$. The parameters are chosen as $%
\protect\delta=0.001$ and $\protect\epsilon^{\prime }=0.01$. The bound $kf(%
\mathcal{L}_m-\protect\epsilon)$ depends on the input probability
distribution $P(xyz)$ through the parameter $r=\min_{xyz}P(xyz)$. The
blue-square line represents the bound under a uniform distribution ($%
P(xyz)=1/4$ for all $(x,y,z)\in \mathcal{S}$), while the red-dotted line
shows the bound under a biased probability distribution with $%
P(011)=P(101)=P(110)=\protect\alpha k^{-1/2}$ and $P(000)=1-3\protect\alpha %
k^{-1/2}$ with $\protect\alpha=10$. It consumes less randomness to generate
a biased distribution for the input bits, so the net amount of randomness,
defined as the number of output random bits minus that of the input, becomes
positive when $k$ is large (typically $k$ needs to be of the order $10^5$).
The inset (b) plots the net amount of randomness generated after $k$ trails
under a biased distribution of the inputs. The parameters are the same as
those in the inset (a). }
\label{Randomness-Show}
\end{figure}

We need to show that the output string $\mathcal{O}$ from the measurement
outcomes contains genuine randomness by proving that it has a nonzero
entropy. Let $\{\mathcal{L}_{m}:0\leq m\leq m_{max}\}$ be a series of
violation thresholds with $\mathcal{L}_{0}=2$ and $\mathcal{L}_{m_{max}}=4$,
corresponding respectively to the classical and quantum bound. Denote by $%
\mathcal{D}(m)$ the probability that the observed violation $\hat{L}$ lies
in the interval $[\mathcal{L}_{m},\mathcal{L}_{m+1})$. We can use the
min-entropy to quantify randomness of the output string $\mathcal{O}$~\cite%
{2010Pironio,2012Pironio,2009Koenig}:
\begin{equation}
E_{\infty }(\mathcal{O}|\mathcal{I},\mathcal{E},m)_{\mathcal{D}}\equiv -%
\mathtt{log}_{2}\sum_{\mathcal{I},\mathcal{E}}[\max_{\mathcal{O}}\mathcal{D}(%
\mathcal{O},\mathcal{I},\mathcal{E}|m)],
\end{equation}%
where $\mathcal{E}$ represents the knowledge that a possible adversary has
on the state of the device and the maximum is taken over all possible values
of the output string $\mathcal{O}$. The probability distribution $\mathcal{D}%
(\mathcal{O},\mathcal{I},\mathcal{E}|m)$ is defined in the Supplemental
Material. Based on a similar procedure as in Ref.~\cite{2010Pironio}, we
can prove that if $\mathcal{D}(m)>\delta $, the min-entropy of the output
string conditional on the input string and the adversary's information has a
lower bound (see derivation in the supplement), given by
\begin{equation}
E_{\infty }(\mathcal{O}|\mathcal{I},\mathcal{E},m)_{\mathcal{D}}\geq kf(%
\mathcal{L}_{m}-\epsilon )-\log _{2}\left( 1/\delta \right) ,
\label{min-entropyBound}
\end{equation}%
where the parameter $\epsilon \equiv \sqrt{-2(1+4r)^{2}(\ln \epsilon
^{\prime 2})}$ with $r=\min P(xyz)$, the smallest probability of the input
pairs, and $\epsilon ^{\prime }$ is a given parameter that characterizes the
closeness between the target distribution $\mathcal{D}(\mathcal{O},\mathcal{I%
},\mathcal{E})$ and the real distribution after $k$ successive measurements
(see the supplement for an explicit definition). The function $f(\hat{L})$
can be obtained through numerical calculation based on semi-definite
programming (SDP)~\cite{1996Vandenberghe} and is shown in Fig. \ref%
{Randomness-Show}. The minimum-entropy bound $kf(\mathcal{L}_{m}-\epsilon
)-\log _{2}\frac{1}{\delta }$ and the net entropy versus the number of
trials $k$ are plotted in the insets (a) and (b) of Fig. \ref%
{Randomness-Show}. Any observed quantum violation with $\hat{L}>2$ leads to
a positive lower bound of the min-entropy, and a positive mini-entropy
guarantees that genuine random numbers can be extracted from the string $%
\mathcal{O}$ of the measurement outcomes through the standard protocol of
random number extractors \cite{1999Nisan}. As some amount of randomness
needs to be consumed to prepare the input string according to the
probability distribution $P(xyz)$, the scheme here actually realizes a
randomness expansion device~\cite{2007Colbeck,2010Pironio}. Similar to Ref.
\cite{2010Pironio}, we can show that under a biased distribution $P(xyz)$ as
shown in Fig.~\ref{Randomness-Show} we generate a much longer random output
string of length $O(k)$~from a relatively small amount of random seeds of
length $O(\sqrt{k}\log _{2}\sqrt{k})$ when $k$ is large.

We now show how to generate and certify random numbers using Majorana
fermions. The key step is to generate a three-qubit entangled state and find
suitable measurements that lead to violation of the MABK inequality.
Majorana fermions are non-Abelian anyons, and their braiding gives
nontrivial quantum operations. However, this set of operations are very
restricted. First, all the gates generated by topological manipulation of
Majorana fermions belong to the Clifford group, and it is impossible to use
such operations alone to violate the CHSH inequality \cite{2009Brennen}. We
have to consider instead the multi-qubit MABK inequality. Second, it is not
obvious that one can violate the MABK inequality as well using only
topological operations. There are two ways to encode a qubit using Majorana
fermions, using either two quasiparticles (Majorana fermions) or four
quasiparticles (see the details in the supplement). In the two-quasiparticle
encoding scheme, although the braiding gates exhaust the entire two-qubit
Clifford group, they cannot span the whole Clifford group for more than two
qubits \cite{2009Ahlbrecht}. Furthermore, braiding Majorana fermions within
each qubit cannot change the topological charge of this qubit which fixes
the measurement basis. Thus, no violation of the MABK inequality can be
achieved using the topological operations alone in the two-quasiparticle
encoding scheme. In the four-qusiparticle encoding scheme, it is not
straightforward either as braidings in this scheme only allows certain
single-qubit rotations and no entanglement can be obtained due to the
no-entanglement rule proved already for this encoding scheme~\cite%
{2006Bravyi}.

Fortunately, we can overcome this difficulty by taking advantage of the
non-destructive measurement of the anyon fusion, which can induce qubit
entanglement~\cite{2005Bravyi}. In a real physical device, the anyon fusion
can be read out non-destructively through the anyon interferometry \cite%
{2010Hassler}. In the four-qusiparticle encoding scheme: each qubit is
encoded by four Majorana fermions, with the total topological charge $0$.
The qubit basis-states are represented by $|0\rangle \equiv |((\bullet
,\bullet )_{\mathbf{I}},(\bullet ,\bullet )_{\mathbf{I}})_{\mathbf{I}%
}\rangle $ and $|1\rangle \equiv |((\bullet ,\bullet )_{\psi },(\bullet
,\bullet )_{\psi })_{\mathbf{I}}\rangle $. Here, each $\bullet $ represents
a Majorana fermion; $\mathbf{I}$ and $\psi $ represent the two possible
fusion channels of a pair of Majorana fermions, with $\mathbf{I}$ standing
for the vacuum state and $\psi $ denoting a normal fermion. As explained in
the Supplemental Material, a topologically protected two-qubit CNOT gate
can be implemented using braidings together with non-destructive
measurements of the anyon fusion~\cite{2005Bravyi}. To certify randomness
through the MABK\ inequality, we need to prepare a three-qubit entangled
state. For this purpose, we need in total fourteen Majorana fermions, where
twelve of them are used to encode three qubits and another ancillary pair is
required for implementation of the effective CNOT gates through measurement
of the anyon fusion. Initially, the logical state is $|\Phi \rangle
_{i}=|000\rangle $. We apply first a Hadamard gate on the qubit $1$, which
can be implemented through a series of anyon braiding as shown in Fig.~\ref%
{Braidings}b, and then two effective CNOT gates on the logical qubits $1$, $2
$, and $2,$ $3$. The final state is the standard three-qubit maximally
entangled state $|\Psi \rangle _{f}=(|000\rangle +|111\rangle )/\sqrt{2}$.
After $|\Psi \rangle _{f}$ is generated, the three qubits can be separated
and we need only local braiding and fusion of anyons within each qubit to
perform the measurements in the appropriate bases to generate random numbers
and certify them through test of the MABK\ inequality.

\begin{figure}[tbp]
\includegraphics[width=85mm]{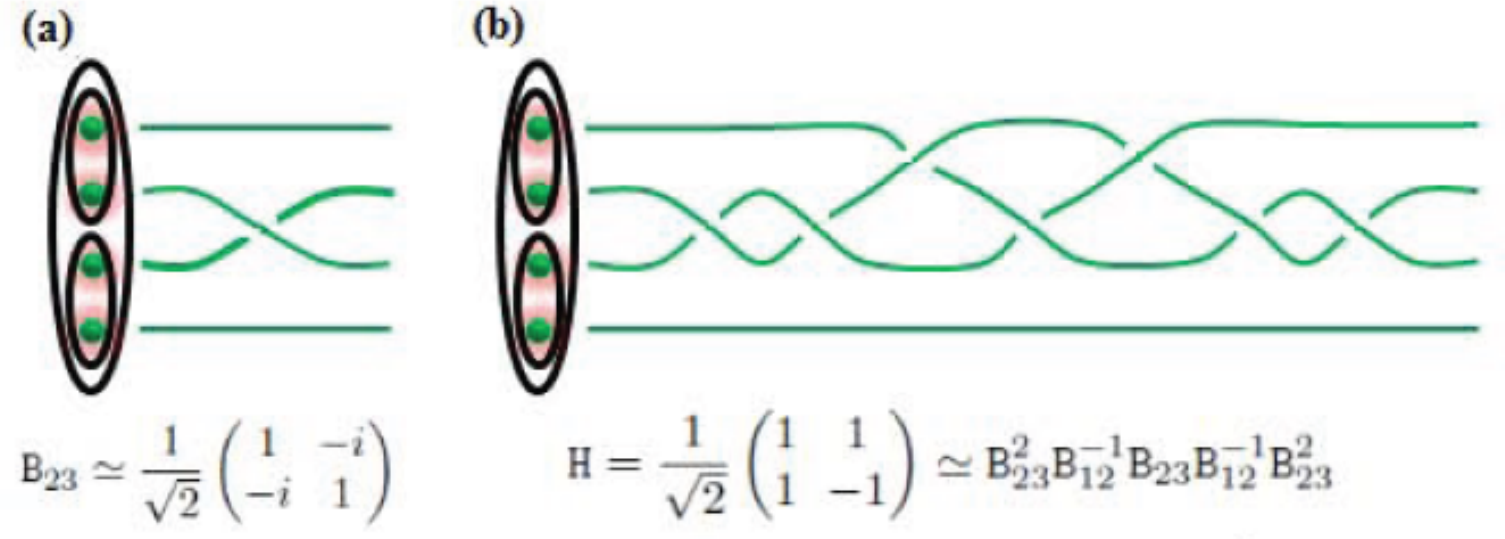}\newline
\caption{(Color online) Illustration of the encoding scheme for a logic
qubit using Majorana fermions and two single-qubit operations that can be
implemented through anyon braiding. Each qubit is encoded by four Majorana
fermions. (a)A counterclockwise braiding of Majorana fermions $2$ and $3$
implements a unitary gate $\mathtt{B}_{23}$ on the corresponding qubit. (b)
Implementation of the Hadamard gate through composition of anyon braiding.
In both (a) and (b), time flows from left to right and $\simeq$ means equal
up to an irrelevant overall phase. }
\label{Braidings}
\end{figure}

To perform the measurements, we read out each qubit according to the input
string $\mathcal{I}$ through nondestructive detection of the anyon fusion.
If the input is $0$, we first braid the Majorana fermions to implement a
Hadamard gate $H$ on this qubit (as shown in Fig.~\ref{Braidings}b), and
then measure the fusion of the first two Majorana fermions within each
qubit. The measurement outcome is $0$ ($1$) if the fusion result is $\mathbf{%
I}$ ($\psi $). If the input is $1$, we first braid the Majorana fermions to
implement a $B_{23}$ gate (see Fig.~\ref{Braidings}a) on this qubit before
the same readout measurement. For instance, with the the input $%
(x,y,z)=(0,1,1)$, we apply a Hadamard gate to the first qubit and $B_{23}$
gates to the second and the third qubits, followed by the nondestructive
measurement of fusion of the first two Majorana fermions in each qubit.
Under the state $|\Psi \rangle _{f}$, the conditional probability of the
measurement outcomes $(a,b,c)$ under the measurement setting $(x,y,z)$ for
these three qubits is give by
\begin{equation}
P(abc|xyz)=\left\vert \langle abc|(\mathtt{U}_{x}\mathtt{U}_{y}\mathtt{U}%
_{z})|\Psi \rangle _{f}\right\vert ^{2},
\end{equation}%
where $\mathtt{U}_{0}=H$ and $\mathtt{U}_{1}=B_{23}$. With this conditional
probability, we find the expected value of $\hat{L}$ defined in Eqs. (1,2)
is $\hat{L}=4$, achieving the maximum quantum violation of the MABK\
inequality. All the steps for measurements and state preparation are based
on the topologically protected operations such as anyon braiding or
nondestructive detection of the anyon fusion, so the scheme here is
intrinsically fault-tolerant and we should get the ideal value of $\hat{L}=4$
if the Majorana fermions can be manipulated at will in experiments. Such a
large violation perfectly certifies genuine randomness of the measurement
outcomes.

In summary, we have shown that genuine number numbers can be generated and
certified through topologically manipulation of Majorana fermions, a kind of
anyonic excitations in engineered materials. Such a protocol is
intrinsically fault-tolerant. Given the rapid experimental progress on
realization of Majorana fermions in real materials \cite{2012Mourik,2012Wang}%
, this protocol offers a promising prospective for application of these
topological particles in an important direction of cryptography with broad
implications in science and engineering.

We thank Y. H. Chan, J. X. Gong, and E. Lichko for
discussions. This work was supported by the NBRPC (973 Program) 2011CBA00300
(2011CBA00302), the IARPA MUSIQC program, the ARO and the AFOSR MURI program.

\begin{widetext}
\section{Supplementary information: Fault Tolerant Quantum Random Number Generator Certified by Majorana
Fermions}
This supplementary information gives more details about realization of
fault-tolerant quantum random number generator through topological
manipulation of Majorana fermions. In Sec. I, we give the detailed proof on
how to certify genuine randomness through observation of violation of the
MABK\ inequality. In Sec. II, we summarize the topological properties of
Majorana fermions and show the implementation of the necessary topological
quantum gates on the logic qubits encoded with these Majorana fermions.

\subsection{Randomness certified by observation of violation of the MABK
inequality}

In this section, we establish a link between randomness of the measurement
outputs of a quantum system and violation of the MABK inequality. A link
between randomness and violation of the Bell-CHSH inequality has been
established in Ref.~\cite{2010Pironio,correction}. Here, we generalize the
result from the two-qubit CHSH inequality to the three-qubit MABK\
inequality. Consider a quantum nonlocality test on three qubits. Each qubit
has two settings of two-outcome measurements, denoted by $\left\{
x,y,z\right\} $, respectively for the three qubits. The measurement outputs $%
\left\{ a,b,c\right\} $ of this quantum system are characterized by the
joint probability distribution $P=\{P(abc|xyz)\}$. Randomness of the outputs
$\left\{ a,b,c\right\} $ are quantified by the min-entropy, defined as $%
E_{\infty }(ABC|XYZ)=-\mathtt{log}_{2}[\max_{abc}P(abc|xyz)]$. With an
experimental observation of violation $\hat{L}$ of the MABK inequality, our
aim is to find a lower bound on the min-entropy
\begin{equation}
E_{\infty }(ABC|XYZ)\geq f(\hat{L}).  \label{siglef}
\end{equation}%
This is equivalent to solving of the following optimization problem~\cite%
{2010Pironio}: \
\begin{eqnarray}
P^{\ast }(abc|xyz)=\; &\max &\;P(abc|xyz)  \notag  \label{Min-entropy-Single}
\\
&\mathtt{subject}\text{ }\mathtt{to}&\;L=\hat{L} \\
&&P(abc|xyz)=\mathtt{Tr}(\rho M_{x}^{a}\otimes M_{y}^{b}\otimes M_{z}^{c})
\notag
\end{eqnarray}%
where $L$ is defined in Eq.(2) of the main text and $(\rho
,M_{x}^{a},M_{y}^{b},M_{z}^{c})$ constitutes a quantum realization of the
Bell scenario~\cite{2011Acin}. Thus, the minimal value of $E_{\infty
}(ABC|XYZ)$ compatible with the MABK violation $\hat{L}$ and quantum theory
is given by $E_{\infty }(ABC|XYZ)=-\mathtt{log}_{2}[\max_{abc}P^{\ast
}(abc|xyz)]$. Consequently, to obtain $f(\hat{L})$ we only need to solve (%
\ref{Min-entropy-Single}) for all possible input and output triplets $(x,y,z)
$ and $(a,b,c)$. This can be effectively done by casting it to a \textit{%
semi-definite program} (SDP)~\cite{1996Vandenberghe}. An infinite hierarchy
of conditions that need to be satisfied by all quantum correlations are
introduced in Ref.~\cite{2007Navascues,2008Navascues,2009Pironio}. All these
conditions can be transformed to a SDP problem and the hierarchy is complete
in the asymptotic limit, i.e., it guarantees existence of a quantum
realization if all the conditions in the hierarchy are satisfied. Generally,
conditions higher in the hierarchy are more constraining and thus better
reflect the constraints in (\ref{Min-entropy-Single}) and give a tighter
lower bound. To obtain a lower bound of the min-entropy for a given MABK
violation $\hat{L}$, we use the matlab toolbox SeDuMi~\cite{Sturm-Sedumi}
and solve the SDP corresponding to the certificates between order $1$ and
order $2$~\cite{2007Navascues}. The result is plotted in Fig.1 in the main
text. From the figure, $f(\hat{L})$ equals zero at the classical point $\hat{%
L}=2$ and increases monotonously as the MABK violation $\hat{L}$ increases.
For the maximal violation $\hat{L}=4$, $P^{\ast }\approx 0.5003$,
corresponding to $f(\hat{L})\simeq 0.9991$ bits.

Equation (4) in the main text can be derived using arguments similar to
those in Ref.~\cite{2010Pironio,2012Pironio}. The difference is that
the Bell scenario in Refs.~\cite{2010Pironio} is based on the two-qubit CHSH inequality, which
needs to be extended in our scheme with the three-qubit MABK inequality. Suppose we run the
experiments $k$ times and denote the input and output string as $\mathcal{I}%
=(x_{1},y_{1},z_{1};\cdots ;x_{k},y_{k},z_{k})$ and $\mathcal{O}%
=(a_{1},b_{1},c_{1};\cdots ;a_{k},b_{k},c_{k})$, respectively. As in the
main text, let $\{\mathcal{L}_{m}:0\leq m\leq m_{max}\}$ be a series of MABK
violation thresholds, and denote $\mathcal{D}(m)$ the probability that the
observed KCBS violation $\hat{L}$ lies in the interval $[\mathcal{L}_{m},%
\mathcal{L}_{m+1})$. Denote by $\mathcal{E}$ the possible classical side
information an adversary may have. To derive Eq. (4) in the main text, let
us first introduce the following theorem:

\textbf{Theorem 1}. Suppose the experiments are carried out $k$ times and
each triplet of inputs $(x_{i},y_{i},z_{i})$ is generated independently with
probability $P(xyz)$. Let $\delta $, $\epsilon ^{\prime }>0$ be two
arbitrary parameters and $r=\min \{P(xyz))\}$, then the distribution $P(%
\mathcal{O}\mathcal{I}\mathcal{E})$ characterizing $k$ successive use of the
devices is $\epsilon ^{\prime }$-close to a distribution $\mathcal{D}$ such
that, either $\mathcal{D}(m)\leq \delta $ or
\begin{equation}
E_{\infty }(\mathcal{O}|\mathcal{I},\mathcal{E},m)_{\mathcal{D}}\geq kf(%
\mathcal{L}_{m}-\epsilon )+\log _{2}\delta ,  \label{theorem1bound}
\end{equation}%
where $\epsilon =(4+1/r)\sqrt{-2\ln \epsilon ^{\prime }/k}$.

Equation~(\ref{theorem1bound}) is equivalent to Eq. (4) in the main text.
Theorem 1 tells us that the distribution $P$, which characterizes the output
$\mathcal{O}$ of the device and its correlation with the input $\mathcal{I}$
and the adversary's classical side information $\mathcal{E}$, is basically
indistinguishable from a distribution $\mathcal{D}$ that will be defined
below~\cite{2012Pironio}. If we find that the observed MABK violation $\hat{L%
}$ lies in $[\mathcal{L}_{m},\mathcal{L}_{m+1})$ with a non-negligible
probability, i.e., $\mathcal{D}(m)>\delta $, the entropy of the outputs $%
\mathcal{O}$ is guaranteed to have a positive lower bound $kf(\mathcal{L}%
_{m}-\epsilon )-\log _{2}\frac{1}{\delta }$, that is, the randomness of the
outputs is guaranteed to be larger than $kf(\mathcal{L}_{m})$ up to
epsilonic correction.

\textit{Proof}. We use a procedure similar to those in Ref.~\cite%
{2012Pironio} to prove the above theorem. Let us define a function $\mathcal{%
G}(L)=2^{-f(L)}$, which is concave and monotonically decreasing given by the
solution of the optimization problem in Eq.~(\ref{Min-entropy-Single})
(shown in Fig. 1 of the main text). Denote by $\mathcal{O}%
^{n}=(a_{i},b_{i},c_{i};\cdots ;a_{n},b_{n},c_{n})$ ($n\leq k$) the string
of outputs before the $(n+1)$th round of experiment (similarly, $\mathcal{I}%
^{n}$ denotes the string of inputs). We introduce an indicator function $%
\chi (e)$ as: $\chi (e)=1$ if the event $e$ happens and $\chi (e)=0$
otherwise. Consider the following random variable
\begin{equation}
\hat{L}_{i}=\sum_{abc;(x,y,z)\in \mathcal{S}}\tau (x,y,z)\Lambda (a,b,c)%
\frac{\chi (a_{i}=a,b_{i}=b,c_{i}=c;x_{i}=x,y_{i}=y,z_{i}=z)}{P(xyz)},
\label{randomvar}
\end{equation}%
where $\mathcal{S}$ and $\tau (x,y,z)$ are defined in the main text, and $%
\Lambda (a,b,c)=1$ if $a+b+c$ is $\mathtt{even}$ and $\Lambda (a,b,c)=-1$ if
$a+b+c$ is $\mathtt{odd}$. It is easy to check that Eq.(\ref{randomvar})
reduces to the MABK expression (2) in the main text and the expectation
value of $\hat{L}_{i}$ conditional on the past $W^{i}$ is equal to $L(W^{i})$%
, i.e., $\mathbb{E}(\hat{L}_{i}|W^{i})=L(W^{i})$. We use $\mathcal{W}%
^{i}\equiv (\mathcal{O}^{i-1}\mathcal{I}^{i-1}\mathcal{E})$ to denote all
the events before the $i$th round of experiment and the possible adversary's
classical side information. The estimator of the MABK violation can be
defined as: $\hat{L}=\frac{1}{k}\sum_{i=1}^{k}\hat{L}_{i}$. With these
notations, first we introduce two lemmas for proof of the main theorem.

\textbf{Lemma 1}. For any given parameter $\epsilon^{\prime }>0$, let $%
\epsilon=(4+1/r)\sqrt{-2\ln\epsilon^{\prime }/k}$ and $S_{\epsilon}=\{(%
\mathcal{O},\mathcal{I},\mathcal{E})|\frac{1} {k}\sum_{i=1}^k\mathbb{E}(\hat{%
L}_i|W^i)\geq \hat{L}(\mathcal{O},\mathcal{I})-\epsilon\}$, then we have:

(i) for any $(\mathcal{O},\mathcal{I},\mathcal{E})\in S_{\epsilon }$,
\begin{equation}
P(\mathcal{O}|\mathcal{I}\mathcal{E})\leq \mathcal{G}^{k}(\hat{L}(\mathcal{O}%
,\mathcal{I})-\epsilon ).  \label{POIE}
\end{equation}

(ii)
\begin{equation}
\mathtt{Pr}(S_{\epsilon })=\sum_{(\mathcal{O},\mathcal{I},\mathcal{E})\in
S_{\epsilon }}P(\mathcal{O},\mathcal{I},\mathcal{E})\geq 1-\epsilon ^{\prime
}.  \label{PrT1}
\end{equation}

\textit{Proof}. According to the Bayes' rule and the fact that the response
of a system does not depend on the future inputs and outputs, we have:

\begin{eqnarray}
P(\mathcal{O}|\mathcal{I}\mathcal{E}) &=&\prod_{i=1}^{k}P(a_{i}b_{i}c_{i}|%
\mathcal{O}^{i-1}\mathcal{I}^{i}\mathcal{E})  \notag  \label{IntroWi} \\
&=&\prod_{i=1}^{k}P(a_{i}b_{i}c_{i}|x_{i}y_{i}z_{i}\mathcal{W}^{i})
\end{eqnarray}%
From the solution to the optimization problem in Eq. (\ref%
{Min-entropy-Single}), the probability $P(a_{i}b_{i}c_{i}|x_{i}y_{i}z_{i}%
\mathcal{W}^{i})$ is bounded by a function of the MABK violation $L(W^{i})$:
$P(a_{i}b_{i}c_{i}|x_{i}y_{i}z_{i}\mathcal{W}^{i})\leq \mathcal{G}(L(W^{i}))$%
. Thus, we have:
\begin{eqnarray}
P(\mathcal{O}|\mathcal{I}\mathcal{E}) &\leq &\prod_{i=1}^{k}\mathcal{G}%
(L(W^{i}))  \notag \\
&\leq &\mathcal{G}^{k}(\frac{1}{k}\mathbb{E}(\hat{L}_{i}|W^{i}))  \notag \\
&\leq &\mathcal{G}^{k}(\hat{L}(\mathcal{O},\mathcal{I})-\epsilon ).
\end{eqnarray}%
Here, to obtain the second inequality, we have used the equality $\mathbb{E}(%
\hat{L}_{i}|W^{i})=L(W^{i})$ and the fact that $\mathcal{G}$ is
logarithmically concave and monotonically decreasing. The third inequality
is obtained from the definition of $S_{\epsilon }$ and the fact that $%
\mathcal{G}$ is decreasing. To get Eq. (\ref{PrT1}), we can define another
random variable $M^{q}=\sum_{i=1}^{q}(\hat{L}_{i}-\mathbb{E}(\hat{L}%
_{i}|W^{i}))$. Then it is easy to verify that (i) $|M^{q}|\leq 2q/r<\infty $%
, (ii) $|\hat{L}_{i}-L(W^{i})|\leq |\hat{L}_{i}|+|L(W^{i})|\leq \frac{1}{r}+4
$, and (iii) $\mathbb{E}(M^{q+1}|W^{q})=M^{q}$. Thus, the sequence $%
\{M^{q}:q\geq 1\}$ is a martingale process~\cite{2001Grimmett-Book}.
Applying the Azuma-Hoeffding inequality $P(M^{q}\geq k\epsilon )\leq \exp (-%
\frac{(k\epsilon )^{2}}{2k(1/r+4)^{2}})$~\cite%
{1967Azuma,1960Hoeffding,2001Grimmett-Book}, we have
\begin{equation}
P\left( \frac{1}{k}\sum_{i=1}^{k}\mathbb{E}(\hat{L}|W^{i})\leq \frac{1}{k}%
\sum_{i=1}^{k}\hat{L}_{i}-\epsilon \right) \leq \epsilon ^{\prime },
\label{MargEq1}
\end{equation}%
where $\epsilon =(4+1/r)\sqrt{-2\ln \epsilon ^{\prime }/k}$. Equation (\ref%
{MargEq1}) combined with the definition of $S_{\epsilon }$ gives Eq.~(\ref%
{PrT1}). Lemma 1 is thus proved.

In the above proof, we only considered the case that the random variable
sequence $\mathcal{O}$ takes values in the output space $\mathbb{S}%
^{k}=\{-1,1\}^{k}$. As in Ref.~\cite{2012Pironio}, we can extend the range
of $\mathcal{O}$ to include \textquotedblleft abort-output" $\bot $, and
view $\mathcal{O}$ as an element of $\mathbb{S}^{k}\cup \bot $ with $P(%
\mathcal{O}|\mathcal{I}\mathcal{E})=0$ if $\mathcal{O}=\bot $. The physical
meaning of $\bot $ is that when $\bot $ is produced by the device, then no
MABK violation has been obtained and no randomness is certified.

\textbf{Lemma 2}. There exists a probability distribution $\mathcal{D}=\{%
\mathcal{D}(\mathcal{O},\mathcal{I},\mathcal{E})\}$, which is $\epsilon
^{\prime }$-close to $P=\{P(\mathcal{O},\mathcal{I},\mathcal{E})\}$, i.e., $%
d(\mathcal{D},P)=\frac{1}{2}\sum_{\mathcal{O},\mathcal{I},\mathcal{E}}|P(%
\mathcal{O},\mathcal{I},\mathcal{E})-\mathcal{D}(\mathcal{O},\mathcal{I},%
\mathcal{E})|\leq \epsilon ^{\prime }$, and satisfies the following
condition
\begin{equation}
\mathcal{D}(\mathcal{O}|\mathcal{I},\mathcal{E})\leq \mathcal{G}^{k}(\hat{L}(%
\mathcal{O},\mathcal{I})-\epsilon ),  \label{Dcondition2}
\end{equation}%
for all $(\mathcal{O},\mathcal{I},\mathcal{E})$ such that $\mathcal{O}\neq
\bot $.

\textit{Proof}. We show how to construct a probability distribution
satisfying the above two conditions. To this end, we introduce $\mathcal{D}(%
\mathcal{O},\mathcal{I},\mathcal{E})=P(\mathcal{I})P(\mathcal{E})\mathcal{D}(%
\mathcal{O}|\mathcal{I},\mathcal{E})$. $\mathcal{D}(\mathcal{O}|\mathcal{I},%
\mathcal{E})$ is defined as:
\begin{equation}
\mathcal{D}(\mathcal{O}|\mathcal{I},\mathcal{E})=\left\{
\begin{array}{cc}
P(\mathcal{O}|\mathcal{I},\mathcal{E}), & \mathtt{if}\;(\mathcal{O},\mathcal{%
I},\mathcal{E})\in S_{\epsilon } \\
0, & \quad \quad \mathtt{if}\;\mathcal{O}\neq \bot \;\mathtt{and}\;(\mathcal{%
O},\mathcal{I},\mathcal{E})\notin S_{\epsilon } \\
1-\sum_{(\mathcal{O},\mathcal{I},\mathcal{E})\notin S_{\epsilon }}P(\mathcal{%
O}|\mathcal{I},\mathcal{E}) & \;\mathtt{otherwise}%
\end{array}%
\right.
\end{equation}%
Then by Lemma 1, it is straightforward to get that the distribution $%
\mathcal{D}$ satisfies Eq.~(\ref{Dcondition2}) for all $(\mathcal{O},%
\mathcal{I},\mathcal{E})$ with $\mathcal{O}\neq \bot $. The distance between
$P$ and $\mathcal{D}$ can be calculated as:
\begin{eqnarray}
d(\mathcal{D},P) &=&\frac{1}{2}\sum_{\mathcal{O},\mathcal{I},\mathcal{E}}|P(%
\mathcal{O},\mathcal{I},\mathcal{E})-\mathcal{D}(\mathcal{O},\mathcal{I},%
\mathcal{E})|  \notag \\
&=&\frac{1}{2}\sum_{\mathcal{I},\mathcal{E}}P(\mathcal{I},\mathcal{E})\sum_{%
\mathcal{O}}|P(\mathcal{O}|\mathcal{I},\mathcal{E})-\mathcal{D}(\mathcal{O}|%
\mathcal{I},\mathcal{E})|  \notag \\
&=&\frac{1}{2}[\sum_{(\mathcal{O},\mathcal{I},\mathcal{E})\notin \mathcal{T}%
_{\epsilon }}P(\mathcal{O},\mathcal{I},\mathcal{E})+1-\sum_{(\mathcal{O},%
\mathcal{I},\mathcal{E})\in \mathcal{T}_{\epsilon }}P(\mathcal{O},\mathcal{I}%
,\mathcal{E})] \\
&\leq &\epsilon ^{\prime }.  \notag
\end{eqnarray}%
This proves Lemma 2.

With Lemma 2, now the proof of Theorem 1 becomes straightforward. Define a
subset of the outputs as $\mathcal{X}_{m}=\{\mathcal{O}|\mathcal{O}\neq \bot
\;\mathtt{and}\;\mathcal{L}_{m}\leq \hat{L}<\mathcal{L}_{m+1}\}$ and let
$\mathcal{D}(\mathcal{O},\mathcal{I},
\mathcal{E}|m)$ denote the distribution of $\mathcal{O},\mathcal{I},
\mathcal{E}$ conditioned on a particular value of $m$, then we
have:
\begin{eqnarray}
E_{\infty }(\mathcal{O}|\mathcal{I},\mathcal{E},m)_{\mathcal{D}}&\equiv& -%
\mathtt{log}_{2}\sum_{\mathcal{I},\mathcal{E}}[\max_{\mathcal{O}}\mathcal{D}(%
\mathcal{O},\mathcal{I},\mathcal{E}|m)]\nonumber\\
 &= &-\mathtt{log}_{2}\sum_{\mathcal{I},\mathcal{E}}\mathcal{D}(\mathcal{I},%
\mathcal{E}|m)[\max_{\mathcal{O}}\mathcal{D}(\mathcal{O}|\mathcal{I},%
\mathcal{E},m)]  \notag \\
&=&-\mathtt{log}_{2}\sum_{\mathcal{I},\mathcal{E}}\mathcal{D}(\mathcal{I},%
\mathcal{E}|m)\frac{1}{\mathcal{D}(m|\mathcal{I},\mathcal{E})}\max_{\mathcal{%
O}\in \mathcal{X}_{m}}\mathcal{D}(\mathcal{O}|\mathcal{I},\mathcal{E})
\notag \\
&\geq &-\mathtt{log}_{2}\sum_{\mathcal{I},\mathcal{E}}\mathcal{D}(\mathcal{I}%
,\mathcal{E}|m)\frac{\mathcal{G}^{k}(\mathcal{L}_{m}-\epsilon )}{\mathcal{D}%
(m|\mathcal{I},\mathcal{E})} \\
&=&-\mathtt{log}_{2}\sum_{\mathcal{I},\mathcal{E}}\frac{\mathcal{D}(\mathcal{%
I},\mathcal{E})}{\mathcal{D}(m)}\mathcal{G}^{k}(\mathcal{L}_{m}-\epsilon )
\notag \\
&=&kf(\mathcal{L}_{m}-\epsilon )-\log _{2}\frac{1}{\mathcal{D}(m)}.  \notag
\end{eqnarray}%
Here we have used the Bayes' rule in the first, second and the fourth equalities
and Eq.~(\ref{Dcondition2}) from Lemma 2 in the third inequality; for the
last equality, the equation $f=-\log _{2}\mathcal{G}$ is used. The last
equality immediately leads to the claim in Theorem 1.  This concludes the
proof.

It is worthwhile to clarify that in deriving Eq.~(\ref{theorem1bound}) we
have made the following four assumptions~\cite{2010Pironio,2012Pironio}: (i)
the system can be described by quantum theory; (ii) the inputs at the $j$th
trial $(x_{j},y_{j},z_{j})$ are chosen randomly and their values are
revealed to the systems only at step $j$; (iii) the three qubits are
separated and non-interacting during each measurement step. (iv) the
possible adversary has only classical side information. There are no
constraints on the states, measurements, or the Hilbert space. Moreover,
there is even no requirement that the system behaves identically and
independently for each trial. In particular, the system could have an
internal memory (classical or quantum) so that the results of the $j$th
trial depend on the previous $j-1$ trials.

We also note that there is a significant difference between the two-qubit
scenario in Ref. \cite{2010Pironio} and our three-qubit scenario here. In
the two-qubit case, the randomness can be certified by the no-signalling
conditions as well without the assumption of  quantum mechanics. However, in
our three-qubit scenario, the no-signalling conditions are not sufficient to
certify randomness. Actually, we have numerically checked that even for the
maximal possible MABK violation $\hat{L}_{max}=4$, $P^{\ast }(abc|xyz)$ can
be equal to the unity for certain $(a,b,c)$ and $(x,y,z)$ if only the
no-signalling conditions are imposed, which cannot certify any randomness. A
possible reason for this difference is that the MABK inequality only
contains four out of eight possible correlations. In other words, the input
choices $\mathcal{S}$ is only a subset of $\{(x,y,z)|x,y,z=0,1\}$. As a
result, the no-signalling constraints become less effective.

\subsection{Encoding and operation of qubits by topological manipulation of
Majorana fermions }

In this section, we discuss in detail how to control the logical qubits
encoded with Majorana fermions. The fusion rule of Majorana fermions is of
the Ising type: $\tau \times \tau \sim \mathbf{I}+\psi $, where $\tau $, $%
\mathbf{I}$, and $\psi $ stand for a Majorana fermion, the vacuum state, and
a normal fermion, respectively. Generally, there are two encoding schemes.
The first scheme encodes each logical qubit into a pair of Majorana fermions
(two-quasiparticle encoding). When the pair fuse to a vacuum state $\mathbf{I%
}$, we say that the qubit is in state $|0\rangle $; and when they fuse to $%
\psi $, the state is $|1\rangle $. There is also an ancillary pair, which
soak up the extra $\psi $ if necessary to maintain the constraint that the
total topological charge must be $0$ for the entire system~\cite%
{2006Georgiev,2009Ahlbrecht}. In this encoding scheme, braiding operations
of Majorana fermions exhaust the entire two-qubit Clifford group. However,
for three or more qubits, not all Clifford gates could be implemented by
braiding. The embedding of the two-qubit SWAP gate into a $n$-qubit system
cannot be implemented by braiding~\cite{2009Ahlbrecht}. In the
two-quasiparticle encoding scheme, no violation of the MABK inequality can
be obtained as we cannot change the measurement basis through local braiding
of Majorana fermions within each logic qubit.

As we mentioned in the main text, we use the four-quasiparticle encoding
scheme where the qubit basis-states are represented by $|0\rangle
=|((\bullet ,\bullet )_{\mathbf{I}},(\bullet ,\bullet )_{\mathbf{I}})_{%
\mathbf{I}}\rangle $ and $|1\rangle =|((\bullet ,\bullet )_{\psi },(\bullet
,\bullet )_{\psi })_{\mathbf{I}}\rangle $.  Let us first consider braiding
operations of Majorana fermions within each logic qubit. Consider four
Majorana operators $c_{i}$ $(i=1,2,3,4)$ in one logic qubit, which satisfy $%
c_{i}^{\dagger }=c_{i}$, $c_{i}^{2}=1$ and the anti-commutation relation $%
\{c_{i},c_{j}\}=2\delta _{ij}$. The Pauli operators in the computational
basis can be expressed as~\cite{2010Hassler}:
\begin{equation}
\sigma ^{x}=-ic_{2}c_{3},\quad \sigma ^{y}=-ic_{1}c_{3},\quad \sigma
^{z}=-ic_{1}c_{2}.
\end{equation}%
Unitary operations can be implemented by counterclockwise exchange of two
Majorana fermions $j<j^{\prime }$:
\begin{equation}
\mathtt{B}_{jj^{\prime }}=e^{(i\pi /4)(ic_{j}c_{j^{\prime }})}.
\end{equation}%
Specifically, we can write down the three basic braiding operators in the
computational basis:
\begin{equation}
\mathtt{B}_{12}=\mathtt{B}_{34}\simeq \left(
\begin{matrix}
1 & 0 \\
0 & i%
\end{matrix}%
\right) ,\;\mathtt{B}_{23}\simeq \frac{1}{\sqrt{2}}\left(
\begin{matrix}
1 & -i \\
-i & 1%
\end{matrix}%
\right) ,
\end{equation}%
where $\simeq $ means that we ignore an unimportant overall phase. Using
these basic braiding operators, a single-qubit Hadamard gate can be
implemented as $\mathtt{H}=\frac{1}{\sqrt{2}}\left(
\begin{matrix}
1 & 1 \\
1 & -1%
\end{matrix}%
\right) \simeq \mathtt{B}_{23}^{2}\mathtt{B}_{12}^{-1}\mathtt{B}_{23}\mathtt{%
B}_{12}^{-1}\mathtt{B}_{23}^{2}$. The corresponding braidings are shown in
Fig.2 of the main text. Note that the set of operations implemented through
composition of $\mathtt{B}_{12}$ and $\mathtt{B}_{23}$ are still very
limited, however, it is fortunate that $\mathtt{B}_{23}$ and $\mathtt{H}$
give all the gates that we need for change of the measurement bases in test
of the MABK\ inequality. As shown in the main text, we actually get maximum
quantum violation of the MABK inequality by randomly choosing either a $%
\mathtt{B}_{23}$ or an $\mathtt{H}$ gate on each logic qubit before
measurement of the anyon fusion.

With only braiding operations of Majorana fermions, no entangling gate can
be achieved for logic qubits in the four-quasiparticle encoding scheme due
to the \textit{no-entanglement rule} proved in Ref.~\cite{2006Bravyi}. In
order to overcome this problem, we need assistance from another kind of
topological manipulation: nondestructive measurement of the anyon fusion,
which can be implemented through the anyon interferometry as proposed in
Ref.~\cite{2010Hassler}. Suppose that we have eight Majorana modes $%
c_{1},c_{2},\ldots ,c_{8}$, where the first (last) four modes encode the
control (target) qubit, respectively. As shown in Ref. \cite%
{2002Bravyi,2005Bravyi}, a two-qubit controlled phase flip gate $\Lambda
(\sigma ^{z})$ can be implemented through the following identity:
\begin{equation}
\Lambda (\sigma ^{z})=e^{-(\pi /4)c_{3}c_{4}}e^{-(\pi /4)c_{5}c_{6}}e^{(i\pi
/4)c_{4}c_{3}c_{5}c_{6}}e^{i\pi /4}.  \label{conZ}
\end{equation}
Note that the first two operations in Eq.~(\ref{conZ}) can be directly
implemented by braiding operations. The key step is to implement the
operation $e^{(i\pi /4)c_{4}c_{3}c_{5}c_{6}}$. To this end, we use another
ancillary pair of Majorana fermions $c_{9}$ and $c_{10}$. We measure fusion
of the four Majorana modes $c_{4}c_{3}c_{6}c_{9}$. The outcome is $\pm 1$,
corresponding to either a vacuum state ($+1$) or a normal fermion ($-1$) .
The corresponding projector is given by $\Pi _{\pm }^{(4)}=\frac{1}{2}(1\pm
c_{4}c_{3}c_{6}c_{9})$. Then, we similarly measure fusion of the Majorana
modes (operator) $-ic_{5}c_{9}$, with the project denoted by $\Pi _{\pm
}^{(2)}=\frac{1}{2}(1\mp ic_{5}c_{9})$ corresponding to the measurement
outcomes $\pm 1$. We have the following relation~\cite{2002Bravyi,2005Bravyi}:

\begin{eqnarray}
e^{(i\pi /4)c_{4}c_{3}c_{5}c_{6}}=2\dsum\limits_{\eta ,\zeta =\pm }U_{\eta
\zeta }\Pi _{\eta }^{(2)}\Pi _{\zeta }^{(4)},
\end{eqnarray}
where $U_{++}=e^{(\pi /4)c_{5}c_{10}}$, $U_{+-}=ie^{(\pi
/2)c_{4}c_{3}}e^{(\pi /2)c_{5}c_{6}}e^{(\pi /4)c_{5}c_{10}}$, $%
U_{-+}=ie^{(\pi /2)c_{4}c_{3}}e^{(\pi /2)c_{5}c_{6}}e^{-(\pi /4)c_{5}c_{10}}$%
, and $U_{--}=e^{-(\pi /4)c_{5}c_{10}}$. All the gates $U_{\eta \zeta }$ can
be implemented through one or several braiding operations of Majorana
fermions. So this identity shows that an effective controlled phase flip
gate can be implemented on logic qubits through a combination of anyon
braiding and measurement of anyon fusion. Depending on the measurement outcomes $(\zeta,\eta)$ of $c_{4}c_{3}c_{6}c_{9}$ and $-ic_{5}c_{9}$, one can always apply a suitable correction operator $U_{\eta \zeta }$ to obtain the desired operation $e^{(i\pi /4)c_{4}c_{3}c_{5}c_{6}}$. With controlled phase flip gates,
one can easily realize quantum controlled-NOT (CNOT) gate with assistance
from the Hadamard operations that can be implemented through the anyon
braiding. With CNOT\ and Hadamard gates, we can then prepare the maximally
entangled three-qubit state as required for test of quantum violation of the
MABK\ inequality.

\end{widetext}

\end{document}